\documentclass[11pt]{article}

\usepackage[margin=1in]{geometry}
\usepackage{amsmath,amssymb}
\usepackage{booktabs}
\usepackage[numbers,sort&compress]{natbib}
\usepackage[colorlinks=true,linkcolor=blue,citecolor=blue,urlcolor=blue]{hyperref}

\title{Doom Researching: A Conceptual Framework for Repetitive AI-Assisted Information Seeking, Cognitive Offloading, and the Illusion of Knowing}

\author{
  Santosh Premi Adhikari
}

\date{}

\begin{document}
\maketitle

\begin{abstract}
Generative artificial intelligence (GenAI) systems such as ChatGPT, Claude, and Gemini have made information seeking faster, more conversational, and more cognitively comfortable. These affordances can support learning and productivity, but they can also encourage a repetitive pattern in which users continue querying AI systems for explanations, summaries, comparisons, plans, and reassurance without converting those interactions into durable understanding, decisions, or finished work. This conceptual paper proposes the term \emph{doom researching} to describe this AI-mediated pattern of repetitive information seeking without proportional synthesis or output. Building on research on doomscrolling, information seeking, cognitive offloading, transactive memory, human-AI interaction, productivity loss, and the illusion of knowing, the paper develops a framework in which fluent AI responses reduce cognitive effort, inflate perceived knowledge, and increase the likelihood of further querying. The framework distinguishes doom researching from doomscrolling, cyberchondria, ordinary research, and productive AI-assisted work. It introduces a formal model of the doom researching loop, a candidate risk index for empirical measurement, and testable propositions for future studies. It then extends the construct through the lens of the extended mind thesis, distinguishing assistive, substitutive, and disruptive forms of cognitive offloading, and connects individual doom researching to the broader literature on AI-driven homogenization and knowledge collapse. The paper argues that doom researching is not simply ``too much AI use'' but a misalignment among inquiry, metacognition, and output. The goal is to provide a vocabulary and research agenda for studying when AI-assisted inquiry becomes a substitute for thinking, synthesis, and action.
\end{abstract}

\noindent\textbf{Keywords:} generative AI, human-AI interaction, information seeking, cognitive offloading, metacognition, illusion of knowing, doomscrolling, extended mind, knowledge collapse, epistemic diversity, AI overreliance

\section{Introduction}

Generative AI tools have changed the practical experience of research. A user can ask a chatbot to summarize a field, compare theories, explain a paper, generate a research plan, criticize a draft, design a study, create a reading list, and reformulate the same ideas in multiple levels of difficulty. This interaction style makes research feel immediate and personal. It also reduces many of the frictions that traditionally forced researchers to slow down: searching through sources, reading full arguments, taking notes, deciding what matters, and writing preliminary synthesis.

The observation motivating this paper is personal. While working on a thesis and exploring unfamiliar research areas, I noticed a recurring pattern: hours spent querying AI systems for explanations, comparisons, and frameworks, accumulating many ideas in conversations, yet producing little written output and retaining less than expected. The feeling of having researched something was vivid; the ability to explain or act on it independently was not. This experience, I suspected, was not unique.

The reduction of friction is valuable. AI-assisted search and writing can help users explore unfamiliar domains, overcome blank-page problems, and receive feedback. However, the same affordances can create a paradox: the user may spend more time researching because research has become easier, more stimulating, and less demanding than producing. In this pattern, the user repeatedly asks AI systems for more explanations or better plans, accumulates a large amount of conversational information, feels increasingly informed, but produces little output and retains little stable understanding.

This paper proposes the term \emph{doom researching} for this pattern. The term is inspired by doomscrolling, which refers to persistent consumption of negative information in digital feeds \citep{sharma2022doomscrolling,satici2023doomscrolling}. Doom researching differs from doomscrolling in three ways. First, it is active rather than primarily feed-based: the user asks questions, requests explanations, and initiates searches. Second, it is often instrumentally framed: the user believes they are preparing for productive action. Third, in the AI context, the interaction is conversational and generative; the system does not merely retrieve information but produces fluent artifacts that can feel like understanding.

The central claim is not that AI-assisted research is harmful in general. Rather, the claim is that a particular configuration of AI use can become maladaptive: repeated inquiry continues after its marginal value has declined, cognitive work is outsourced instead of internalized, and perceived knowledge rises faster than actual competence or output. This pattern deserves a name because it sits at the intersection of several literatures that are usually studied separately: doomscrolling and negative information consumption, information seeking under uncertainty, cognitive offloading, AI overreliance, productivity loss in human-AI interaction, and metacognitive illusions.

This paper makes six contributions. First, it defines doom researching as an AI-mediated information behavior. Second, it differentiates doom researching from neighboring constructs including doomscrolling, doomsurfing, cyberchondria, ordinary literature review, procrastination, and productive AI-assisted research. Third, it proposes a conceptual and mathematical model of the doom researching loop. Fourth, it offers testable propositions, measurement ideas, and design implications for future empirical work. Fifth, it develops a phenomenological analysis of the ``manager-junior illusion''---the felt experience of directing an AI assistant that resembles intellectual work more than it resembles the productive activity it displaces---and grounds this analysis in the extended mind thesis and a taxonomy of assistive, substitutive, and disruptive cognitive offloading. Sixth, it connects individual doom researching to the collective-level literature on AI-driven homogenization of ideas and knowledge collapse, and argues that repetitive AI-mediated inquiry has consequences not only for the individual but for the diversity of the ideas a research community produces.

\section{Background}

\subsection{Doomscrolling and repetitive negative information consumption}

Doomscrolling has been conceptualized as a media habit in which users persistently attend to negative information about crises, disasters, and tragedies in social media feeds \citep{sharma2022doomscrolling}. Empirical work has developed doomscrolling scales and linked the behavior to anxiety, fear of missing out, problematic social media use, psychological distress, and reduced wellbeing \citep{sharma2022doomscrolling,satici2023doomscrolling}. Experimental evidence also suggests that even brief exposure to negative pandemic-related social media content can reduce positive affect and optimism \citep{buchanan2021doomscrolling}.

Doomscrolling is relevant because it shows how information seeking can become self-reinforcing even when the information worsens mood or does not improve action. However, doom researching is not identical to doomscrolling. Doomscrolling is usually associated with negative news feeds and passive consumption. Doom researching can involve neutral or even positive topics: thesis ideas, programming architectures, startup strategies, productivity systems, health optimization, or academic fields. The ``doom'' in doom researching does not necessarily come from negative content. It comes from the loop: the user repeatedly seeks more information while delaying synthesis, decision, or output.

\subsection{Information seeking and uncertainty reduction}

Information seeking is often adaptive. People seek information because it can guide action, improve affect, or satisfy cognitive motives such as curiosity and uncertainty reduction \citep{sharot2020information,kelly2021information}. Sharot and Sunstein propose that people estimate the value of information by considering its likely effects on action, affect, and cognition \citep{sharot2020information}. This model is especially useful for doom researching because AI systems strongly amplify the cognitive value of information: they can always provide another explanation, analogy, framework, or list.

Recent computational work suggests that maladaptive information seeking may arise when the motive to reduce uncertainty outweighs the motive to protect affect \citep{cogliati2024maladaptive} (preprint). In AI-mediated research, a related imbalance may occur: the user overweights uncertainty reduction and underweights output. More information feels valuable because it reduces the discomfort of not knowing, but the reduction is temporary. Each answer reveals new concepts, caveats, and possible directions, creating new uncertainty and inviting the next prompt.

\subsection{Cognitive offloading and transactive memory}

Cognitive offloading refers to the use of external actions or tools to reduce internal cognitive demand \citep{riskogilbert2016offloading}. It is not inherently negative. Writing notes, setting reminders, using calculators, and searching the web are all forms of offloading that can support cognition. The problem arises when offloading becomes miscalibrated: the user outsources tasks that are necessary for learning, judgment, or synthesis.

The internet already functions as a form of external or transactive memory. Sparrow, Liu, and Wegner found that when people expect future access to information, they may remember where to find it better than the information itself \citep{sparrow2011google}. Fisher, Goddu, and Keil showed that searching the internet can inflate self-assessed internal knowledge, leading people to mistake access to information for knowledge ``in the head'' \citep{fisher2015internet}. GenAI may intensify this effect because it does not merely point to information; it converts it into fluent, coherent language that resembles the user's own reasoning.

\subsection{Generative AI, metacognition, and productivity loss}

GenAI systems impose new metacognitive demands on users: users must decide what to ask, whether the output is adequate, how much to trust it, when to stop, and how to integrate it into a workflow \citep{tankelevitch2024metacognitive}. These demands are difficult because AI outputs often appear confident and complete even when they are incomplete, biased, or unsupported. Recent human-AI interaction research argues that GenAI can cause productivity loss through role shifts from production to evaluation, interruptions, workflow restructuring, and the tendency of automation to make easy tasks easier while making hard tasks harder \citep{simkute2024ironies}.

Studies of AI-supported reasoning also suggest a disconnect between performance and metacognition. Users may perform better with AI assistance while overestimating their own performance or misunderstanding the source of their success \citep{fernandes2024metacognition}. Work on AI tools and critical thinking similarly raises concerns that frequent AI use may be associated with cognitive offloading and reduced critical engagement \citep{gerlich2025aitools}. Doom researching is one possible behavioral expression of these broader issues.

\section{Defining Doom Researching}

\subsection{Definition}

This paper defines \emph{doom researching} as:

\begin{quote}
Repetitive AI-assisted information seeking in which a user continues querying one or more AI systems for explanations, summaries, comparisons, plans, or reassurance without converting the interaction into proportional synthesis, retained understanding, decision, or external output.
\end{quote}

This definition has five components.

\begin{enumerate}
  \item \textbf{Repetition.} The behavior involves repeated queries, re-prompts, tool switching, or topic cycling.
  \item \textbf{AI assistance.} The behavior is mediated by generative AI systems that produce conversational or document-like answers.
  \item \textbf{Information seeking.} The user's explicit goal is to know, understand, compare, plan, or reduce uncertainty.
  \item \textbf{Low conversion.} The interaction is not converted into a durable artifact, decision, implementation, or tested understanding.
  \item \textbf{Metacognitive inflation.} The user experiences increased perceived knowledge, readiness, or superiority that exceeds actual competence or output.
\end{enumerate}

\subsection{Boundary conditions}

Doom researching is not the same as extensive research. A scholar who reads many papers, takes notes, writes syntheses, and produces a manuscript is not doom researching simply because the process is long. Nor is doom researching the same as using AI frequently. Frequent AI use can be productive if it is structured around goals, verification, and output.

The key distinction is conversion. Productive research converts information into understanding, decisions, artifacts, or action. Doom researching converts information primarily into more information seeking.

\begin{table}[t]
\centering
\caption{Distinguishing doom researching from related constructs.}
\begin{tabular}{p{0.22\linewidth}p{0.34\linewidth}p{0.34\linewidth}}
\toprule
\textbf{Construct} & \textbf{Primary behavior} & \textbf{Difference from doom researching} \\
\midrule
Doomscrolling & Persistent consumption of negative feed content & More passive and feed-based; usually centered on negative news \\
Doomsurfing & Repeated web browsing for distressing information & Search-based but not necessarily AI-mediated or linked to perceived expertise \\
Cyberchondria & Repetitive online health searching that increases health anxiety & Domain-specific to health anxiety \citep{starcevic2020cyberchondria} \\
Ordinary research & Systematic inquiry leading to synthesis or output & High conversion of information into artifacts or decisions \\
Procrastination & Avoidance of intended work & Doom researching may function as procrastination, but is subjectively experienced as productive inquiry \\
AI overreliance & Excessive dependence on AI output & Broader construct; doom researching is a specific loop of repeated inquiry without output \\
\bottomrule
\end{tabular}
\end{table}

\section{A Conceptual Mechanism}

Doom researching is proposed to arise from the interaction of five mechanisms: uncertainty reduction, fluency, cognitive offloading, output avoidance, and metacognitive inflation.

\subsection{Uncertainty reduction}

Research often begins with uncertainty. AI systems are attractive because they quickly reduce the discomfort of not knowing. However, each answer can reveal new branches: alternative theories, missing references, methodological concerns, or better terminology. The result is a partial reduction in local uncertainty but expansion of global uncertainty. The user feels progress because one question has been answered, yet the space of possible questions grows.

\subsection{Fluency}

AI responses are fluent: they are grammatical, structured, confident, and easy to process. Fluency can be mistaken for truth, depth, or personal understanding. In doom researching, the user may experience the answer's clarity as their own clarity. This connects to the illusion of knowing: external access to information is misread as internal knowledge \citep{fisher2015internet}.

\subsection{Cognitive offloading}

AI systems can perform summarization, comparison, planning, critique, and drafting. These are precisely the operations through which researchers normally build internal understanding. When these operations are repeatedly outsourced, the user may reduce the effort needed to form mental models. Offloading becomes problematic when it replaces rather than supports self-explanation.

\subsection{Output avoidance}

Writing, deciding, building, and submitting are costly because they expose gaps in knowledge. AI-assisted research can provide a socially acceptable avoidance behavior: the user is ``still researching.'' In this sense, doom researching can be procrastination disguised as preparation. The user avoids the discomfort of producing by seeking another answer, framework, or plan.

\subsection{Metacognitive inflation}

As the interaction continues, perceived knowledge may rise faster than actual knowledge. The user has seen many explanations and may be able to recognize terms, but recognition is not the same as recall, transfer, or production. This creates a perceived-actual knowledge gap. The user may feel unusually informed, superior, or ready, while still being unable to explain, apply, or publish independently.

\section{A Formal Model of the Doom Researching Loop}

This section proposes a simple formalization intended to guide measurement and hypothesis generation, not to claim a validated or fitted psychological theory. The variables and parameters described below are theoretical constructs. No parameter values have been estimated empirically. The model is intended as a scaffold for future quantitative work.

\subsection{Continuation probability}

Let $t$ index AI-assisted research turns. At each turn, the user decides whether to continue querying the AI system or switch to synthesis/output. The probability of another query is modeled as:

\begin{equation}
P(Q_{t+1}=1) = \sigma(\beta_0 + \beta_u U_t + \beta_f F_t + \beta_r R_t + \beta_o O_t - \beta_s S_t - \beta_p P_t),
\end{equation}

\noindent where $\sigma(x)=1/(1+e^{-x})$ is the logistic function. $U_t$ is unresolved uncertainty, $F_t$ is response fluency, $R_t$ is reassurance or temporary relief, $O_t$ is cognitive offloading opportunity, $S_t$ is synthesis effort already invested by the user, and $P_t$ is concrete output progress. Doom researching becomes more likely when uncertainty, fluency, reassurance, and offloading opportunity are high, while synthesis and output progress are low.

\subsection{Perceived versus actual knowledge}

Let $K^p_t$ be perceived knowledge and $K^a_t$ be actual knowledge at turn $t$. The framework assumes:

\begin{align}
K^p_{t+1} &= K^p_t + \alpha_f F_t + \alpha_n N_t + \alpha_r R_t - \alpha_d D_t, \\
K^a_{t+1} &= K^a_t + \gamma_s S_t + \gamma_e E_t - \delta_o O_t,
\end{align}

\noindent where $F_t$ is response fluency, $N_t$ is novelty of information encountered, $R_t$ is reassurance, and $D_t$ is discovered ignorance---moments in which the AI response reveals a previously unknown gap in the user's understanding (a caveat, a competing theory, an unfamiliar term). The term $-\alpha_d D_t$ allows perceived knowledge to \emph{decrease} within a session, which is the empirically common case for users early in a learning trajectory. For actual knowledge, $S_t$ is synthesis effort, $E_t$ is effortful retrieval or self-explanation, and $O_t$ is substitutive offloading. The perceived-actual knowledge gap is:

\begin{equation}
G_t = K^p_t - K^a_t.
\end{equation}

\noindent A central prediction is that unstructured AI-assisted research increases $G_t$ more than structured AI-assisted research, especially when users do not engage in self-explanation, note synthesis, or output production.

\subsection{Output conversion ratio}

Doom researching depends not only on how much AI is used but on how little is produced relative to use. Because the natural inputs (queries, time, money) are on different scales, they should not be summed directly. A workable operationalization is to standardize each input on a per-session basis and combine them into a single input index. Let $\tilde{Q}$, $\tilde{T}$, and $\tilde{M}$ be within-user z-scores of queries, time spent, and monetary cost, respectively. Define an inquiry intensity index $I_t = w_q\tilde{Q}_t + w_t\tilde{T}_t + w_m\tilde{M}_t$, where the weights $w_q, w_t, w_m \geq 0$ sum to one and are chosen for the domain (e.g., a coding task may weight time more heavily than money). The output conversion ratio is then:

\begin{equation}
C = \frac{z(A)}{I},
\end{equation}

\noindent where $z(A)$ is the standardized measurable artifact output (words drafted, decisions made, code committed, problems solved). A low value of $C$ indicates that much inquiry is producing little external progress. The variables and weights are intended as a scaffold for measurement, not as fitted quantities.

\subsection{Doom Researching Risk Index}

For empirical work, a candidate risk index can be defined as:

\begin{equation}
DRI = z(Q) + z(T) + z(M) + z(CO) + z(G) - z(A) - z(S),
\end{equation}

\noindent where $CO$ is cognitive offloading, $G$ is the perceived-actual knowledge gap, $A$ is artifact output, and $S$ is synthesis effort. Higher $DRI$ values indicate greater risk of doom researching. Equal weights are used here for illustration only; in an empirical instrument the weights should be derived from factor loadings on validated measures. This index is intentionally modular: researchers can adapt the variables to academic writing, software development, health research, or workplace decision-making.

\section{Propositions for Future Research}

The framework yields several testable propositions.

\begin{description}
  \item[P1: Repetition-output dissociation.] Higher numbers of AI prompts will not necessarily predict better outputs. In unstructured conditions, prompt count may become negatively associated with output completion after a threshold.

  \item[P2: Fluency-knowledge inflation.] Fluent AI explanations will increase perceived understanding more than actual recall or transfer, especially when users are not required to explain the concept in their own words.

  \item[P3: Offloading mediation.] Cognitive offloading will mediate the relationship between AI-assisted research frequency and lower independent synthesis quality.

  \item[P4: Uncertainty loop.] Users with higher intolerance of uncertainty or stronger need for closure will be more likely to continue querying AI systems after sufficient information has been obtained.

  \item[P5: Multi-model escalation.] Access to multiple AI systems will increase doom researching risk when users use model switching as reassurance seeking rather than as verification.

  \item[P6: Structured output protection.] Requirements such as source verification, self-explanation, time boxing, and mandatory artifact production will reduce the perceived-actual knowledge gap.
\end{description}

\section{Measurement and Study Designs}

\subsection{Self-report scale development}

A first empirical step would be to develop a Doom Researching Scale. Items could measure repetition, AI tool switching, lack of output, perceived knowledge, and regret. Candidate items include:

\begin{enumerate}
  \item I keep asking AI tools for more explanations even when I already have enough information to start.
  \item I often feel that I understand a topic after reading AI answers, but struggle to explain it without AI.
  \item I switch between AI tools to feel more certain rather than to verify specific claims.
  \item I spend more time asking AI about work than doing the work itself.
  \item AI research makes me feel productive even when I produce no concrete output.
  \item I delay decisions because I want one more AI-generated comparison, summary, or plan.
  \item After long AI research sessions, I often have many ideas but no finished artifact.
  \item I feel temporarily superior or highly knowledgeable after AI-assisted research sessions.
\end{enumerate}

These items should be tested through exploratory factor analysis, confirmatory factor analysis, reliability testing, and construct validation against measures of AI dependence, cognitive offloading, procrastination, metacognition, and output quality.

\subsection{Experimental design}

A simple experiment could assign participants to one of three conditions: no AI research, structured AI research, and unstructured AI research. Participants would research a topic for a fixed time and produce a short explanatory essay or decision memo. Outcome measures would include prompt count, time spent, self-rated understanding, independent recall, transfer performance, and expert-rated output quality.

The key prediction is not that AI use always lowers performance. The stronger prediction is an interaction: structured AI use may improve output, while unstructured repeated AI use may inflate confidence without proportional improvement in quality or retention.

\subsection{Behavioral trace study}

Researchers could collect opt-in interaction logs from students or knowledge workers. Features would include number of prompts, topic switching, repeated requests for summaries, repeated requests for ``best'' options, copy-paste behavior, source checking, and time to first output. These traces could be correlated with artifact completion and independent explanation tests.

\section{Design Implications}

If doom researching is a real AI-mediated behavior, then design should not only optimize answer quality. It should also help users stop researching and start synthesizing. Possible interventions include:

\begin{enumerate}
  \item \textbf{Output prompts.} After several exploratory turns, the system asks the user to produce a paragraph, decision, outline, or claim in their own words.
  \item \textbf{Synthesis checkpoints.} The system periodically asks: ``What do you now believe, and what evidence supports it?''
  \item \textbf{Diminishing returns indicators.} The interface can show when recent prompts are semantically redundant.
  \item \textbf{Time and cost visibility.} The system can display session duration, number of prompts, and estimated subscription/API cost.
  \item \textbf{Source friction.} For academic tasks, the system can require citations and distinguish verified sources from generated synthesis.
  \item \textbf{AI-free retrieval.} The system can hide the answer and ask the user to recall or apply the idea before continuing.
  \item \textbf{Decision closure.} For planning tasks, the system can ask the user to choose one option and define the next external action.
\end{enumerate}

These interventions should be evaluated carefully. Too much friction could reduce accessibility and creativity. The aim is not to make AI less useful but to preserve the cognitive work that turns information into knowledge.

\section{Outsourcing of Intelligence}

\subsection{The manager-junior illusion}

A user who sits down with a generative AI system to explore a research problem may not feel like a passive reader. The interaction can feel like management. The user defines the task, asks follow-up questions, corrects the model when it drifts, and switches to a different model when dissatisfied. The affective texture of this interaction resembles a manager handing a task to a junior colleague and iterating with them: agenda-setting, direction, evaluation, delivery. This is why AI-assisted research can feel unusually productive even when nothing external gets produced.

The felt experience is partially accurate. The user is directing the interaction. But there is a hidden asymmetry between real management and this simulated version of it.

A real manager, to evaluate a junior's deliverable, needs enough independent competence in the domain to judge whether the output is correct, complete, and appropriate. A manager who cannot make that judgment is not managing; they are approving. This is where doom researching produces its most consequential self-deception. The user retains the feeling of oversight while quietly losing the criterion by which oversight is exercised. The user asks the AI a question; the AI produces a fluent answer; the user reads it, feels that it is good, and moves on. Whether the answer is actually correct, whether it selects the right frame, whether it omits the argument that would falsify it---these are judgments the user may no longer make at the depth the topic requires.

The cognitive offloading taxonomy of \citet{jose2025outsourcing} calls this pattern \emph{disruptive offloading}: delegation of the regulatory and reflective processes themselves, not merely delegation of retrieval or computation. Under disruptive offloading, the internal evaluator that makes management possible is precisely what has been outsourced. The role of manager is preserved as posture; the function of manager is quietly transferred to the model.

This claim should be distinguished from a stronger claim that AI users simply believe themselves to be experts from the beginning of an interaction. There is a tension between the manager illusion and the illusion-of-knowing account developed earlier in the paper: in one account, the user feels expert; in the other, the user delegates because they know they are not. These can be understood as phases of the same loop. Early in a session, the user may not feel expert and therefore asks for assistance. As fluent AI answers accumulate, the user's sense of comprehension rises. Later in the session, the user may feel expert enough to accept new answers without pushing back. The manager who started the session was genuinely trying to evaluate; the manager at the end of the session is signing off. Both roles belong to the same doom researching episode.

The affective consequence of this shift is what keeps the loop running. Sending a question to a capable assistant produces a small, immediate sense of having handled the problem. That feeling is not the same as having solved the problem, but for the purpose of continuing the session it is close enough. This is why an afternoon of doom researching can feel like work: the user has completed many small acts of intellectual management, each with a felt closure, without completing the harder work of writing, deciding, or defending a position independently.

This is not a strong neurological claim; such a claim would require empirical evidence beyond the scope of this paper. It is a phenomenological claim about how the pattern feels from the inside, and about why that feeling makes the pattern hard to notice while it is happening. The gap between effort and output remains invisible until the user tries to explain the material independently, or until the user examines what the session actually produced and finds that the answer is: notes, tabs, and a stronger opinion that remains difficult to justify.

\subsection{Theoretical grounding: the extended mind thesis}

The philosophical foundation for understanding what it means to delegate cognitive work to external tools is the \emph{extended mind thesis}, proposed by Clark and Chalmers in 1998. Their argument is that when an external resource---a notebook, a map, a calculator---is reliably available, consistently used, and functionally integrated with biological cognition, it is legitimate to treat that resource as part of the cognitive system. The mind, on this view, extends into the environment \citep{clark1998extended}.

\citet{clark2025extending} applied this framework to generative AI, arguing that AI systems can, in principle, function as extended cognitive components. He introduced the concept of \emph{extended cognitive hygiene}: the set of critical practices needed to ensure that externally delegated cognition contributes to rather than degrades the internal system. Just as we must be discerning about what we allow into our biological minds, we must be discerning about what cognitive processes we allow AI to substitute for us.

The distinction that matters for doom researching is between \emph{augmentation} and \emph{substitution}. Augmentation extends what the researcher can do while preserving or building the researcher's own capacity. Substitution replaces what the researcher would otherwise do, leaving that capacity unused and potentially atrophied. \citet{jose2025outsourcing} formalize this distinction in a three-level taxonomy:

\begin{description}
  \item[Assistive offloading] aids memory or processing without replacing internal effort. The researcher uses AI to retrieve a reference or convert a unit; the intellectual judgment remains internal.
  \item[Substitutive offloading] replaces discrete cognitive operations. The researcher uses AI to summarize a literature section rather than reading and synthesizing independently. The task is completed efficiently, but the internal representation of the material is thin.
  \item[Disruptive offloading] replaces the regulatory and reflective processes that constitute intellectual agency. The researcher uses AI to determine what to think, what questions matter, and what conclusions to draw. Internal cognitive structure degrades; the dependency deepens.
\end{description}

Doom researching operates primarily in the substitutive and disruptive registers. The researcher is not using AI to do more; they are using AI to feel as if they have done something.

\subsection{Implications for the individual researcher}

The effects of intelligence outsourcing on the individual researcher are double-edged, and their balance depends critically on how the outsourcing is structured.

\paragraph{Potential benefits.}
When used at the assistive level, intelligence outsourcing offers genuine value to researchers. AI tools can retrieve literature across disciplines faster than manual search, identify patterns across large text corpora, and surface connections between ideas that a researcher working in one domain may not encounter. For researchers with limited institutional access, AI tools can democratize access to summarized knowledge from expensive journals. For researchers who are non-native speakers of the language in which they must publish, AI assistance with language revision represents a real equity gain. An empirical study of graduate students using generative AI to write open-ended articles found that the high-performance group showed more diversified behavioral transitions, selective information extraction, and deeper processing of AI output, and that they came into the task with significantly higher prior domain knowledge \citep{wang2025cognitive}. One reading of this pattern is that the tool amplified existing capability rather than substituting for it; the strong readers used AI as an adjunct to competence they already possessed.

\paragraph{Risks and costs.}
The same study found that the low-performance group displayed unbalanced and fragmented cognitive structures and primarily engaged with lower-level cognitive processing when using AI. Rather than integrating AI output into their own knowledge framework, they accepted it as a replacement for one. Consistent with this pattern, \citet{jose2025outsourcing} argue---in a conceptual synthesis of the cognitive-offloading literature---that heavy substitutive and disruptive AI use can undermine self-monitoring and metacognitive accuracy. The student who outsources summary writing does not only save time; they lose the synthesis activity that would have built comprehension. The gain in efficiency is offset by a loss in the capacity that efficiency was meant to serve.

For researchers specifically, the risk is compounded by the evaluative dimension of the role. A researcher's core intellectual function is not just to accumulate information but to \emph{assess} it: to judge what is well-supported, what is contested, what is missing, and what should be done next. If these judgments are themselves delegated to the AI, the researcher loses the very capacity that would let them decide whether the AI's judgment is good. Using AI to decide what is worth knowing therefore undermines the metacognitive foundation of research itself. Doom researching, as a pattern of intensive but unproductive AI use, represents the substitutive and disruptive endpoints of this trajectory: the researcher who has outsourced not just retrieval but judgment.

\subsection{Implications for independent and unusual ideas}

The risks of intelligence outsourcing extend beyond the individual researcher to the character of the ideas they produce. Independent and unusual ideas---the intellectual contributions that advance a field beyond its existing center---require a cognitive origin that differs from the statistical mean of existing knowledge. Large language models are, by construction, optimized to produce outputs near the center of the distribution of their training data. The idea that emerges when you ask an AI to help you think about a problem is likely to be the kind of idea that is already present in the existing literature, weighted by prevalence.

This property has documented consequences for creative and intellectual diversity. \citet{moon2025homogenizing} found that each additional AI-generated essay contributed fewer \emph{new} ideas to a corpus than each additional human-generated essay, and that this diversity gap widened at scale. The individual AI output may be locally coherent and fluent; the collective effect of many researchers using AI assistance is a narrowing of the idea space. The authors introduced the concept of \emph{diversity growth rate} to measure this effect: human writers added diverse ideas faster than AI systems.

For the researcher engaged in doom researching, the implication is specific. By repeatedly querying AI systems about a research problem and orienting their thinking around the AI's framings and answers, the researcher's conceptual space narrows toward the AI's distribution. The unusual hypothesis, the non-obvious connection, the contrarian interpretation---ideas that exist at the periphery of the training distribution---become less likely to arise when the researcher's working model of the field is dominated by AI-central outputs. The researcher may retain the formal freedom to think independently, but the cognitive inputs they have absorbed are drawn from the center of the distribution, and those inputs shape what the researcher subsequently generates.

This is a particularly serious concern for researchers in early-stage work, where the differentiation of a research contribution depends on originality. A conceptual framework that is indistinguishable from the aggregated output of existing AI models is not a new contribution; it is a restatement of the consensus. Doom researching, by generating and consuming large volumes of AI output without productive synthesis, increases the probability that any eventual output will be derivative.

\subsection{The homogenization of collective knowledge}

The individual-level effects described above aggregate into a societal-level epistemological risk. \citet{peterson2024collapse} introduced the concept of \emph{knowledge collapse} to describe the progressive narrowing of the set of ideas in active human circulation when AI systems mediate knowledge access at scale. The mechanism is recursive: LLMs generate output toward the center of their training distribution; widespread use of LLMs causes humans to interact primarily with this center; human-generated content shaped by LLM output is re-ingested into future training data; the distribution narrows further. Each cycle reinforces the previous one \citep{qiu2025lockin}.

\citet{qiu2025lockin} tested this hypothesis using real-world GPT usage data and found ``sudden but sustained drops in diversity after the release of new GPT iterations, consistent with the hypothesized human-AI feedback loop.'' The authors describe this as partial support for the hypothesis---in some periods diversity instead increased---but the analysis indicates that the effect is at least detectable in existing data rather than purely hypothetical.

The implication for the research enterprise is substantial. The traditional function of academic research is to extend, challenge, and diversify the knowledge available to a community. Diversity of ideas---including ideas that are odd, marginal, or initially unpopular---is the raw material from which scientific progress is made. A research community whose individual members all use the same AI systems to generate, frame, and evaluate ideas converges on a smaller set of hypotheses, methods, and conclusions. The peer review system, designed to filter bad ideas from good ones, cannot function as a diversity-preserving mechanism if the pool of submitted ideas has already converged.

Doom researching accelerates this dynamic. A researcher who spends many hours per week generating AI output and absorbing it without critical transformation is not contributing to the periphery of the knowledge distribution; they are deepening their alignment with its center. Their eventual research output---if produced---will be more likely to resemble existing work, less likely to identify genuinely novel problems, and less likely to propose heterodox solutions.

A systematic review and three-level meta-analysis by \citet{derooij2025homogenization}, covering 19 studies and 61 effect sizes, reports ``a small but statistically significant homogenization effect associated with AI use, robust across sensitivity analyses.'' The effect was stronger in semantically constrained ideation tasks than in open-ended divergent-thinking tasks, and the authors note that some homogenization ``may persist beyond co-creative episodes and extend to real-world contexts.'' The effect size is small, and the authors are careful not to claim that individual creativity is destroyed by AI use. What they document is a scale effect: when many users interact with similar systems, the aggregate distribution of ideas becomes more compressed. Research ideation is precisely the kind of semantically constrained cognitive task where that compression is likely to matter.

The picture that emerges is not that AI tools eliminate unique thought. It is that they alter its prior probability. Collective knowledge, in a world of widespread doom researching, becomes \emph{additively convergent rather than divergently cumulative}: researchers add variations on the same central themes rather than building an ever-wider and stranger map of what is possible to know.

\section{Ethical and Educational Implications}

Doom researching has implications for education, professional work, and personal wellbeing. In education, students may use AI tools to generate repeated explanations and study plans while avoiding retrieval practice or original writing. In professional work, employees may spend time generating strategy documents, comparisons, and analyses without making decisions. In personal life, users may purchase multiple AI subscriptions and spend large amounts of time seeking certainty, optimization, or self-improvement without action.

The behavior also raises equity concerns. AI tools can be powerful support systems for users with limited access to tutors, mentors, or writing assistance. Overly broad warnings about AI use may disadvantage users who benefit most from these tools. Doom researching should therefore not be framed as a personal moral failure. It should be framed as a pattern of miscalibrated tool use, shaped both by individual habits and by an economic incentive structure in which unstructured engagement is monetized without being distinguished from productive engagement. This paper focuses on the individual cognitive and behavioral construct; the economic dimension is outside its scope. Mitigation is a shared responsibility between users (habits, self-regulation), educators (training in extended cognitive hygiene, in Andy Clark's phrase \citep{clark2025extending}), and platforms (design features that surface the ratio of inquiry to output).

\section{Limitations}

This paper is conceptual and does not present original empirical data. The proposed construct, model, and risk index require validation. The term doom researching may overlap with existing terms such as doomsurfing, cyberchondria, AI overreliance, information overload, and procrastination. Future work should test whether doom researching has discriminant validity or whether it is best understood as a subtype of broader constructs.

Another limitation is that the framework emphasizes risks. Productive AI-assisted research is common and valuable. A user may ask many questions because the task is complex, because they are learning a new field, or because they are using AI as an accessibility tool. High prompt count alone should not be treated as pathological. The central issue is the relationship among repeated inquiry, cognitive effort, perceived knowledge, and output.

\section{Research Agenda}

Future research should answer five questions.

\begin{enumerate}
  \item Can doom researching be measured reliably as a distinct construct?
  \item What user traits predict vulnerability: anxiety, intolerance of uncertainty, perfectionism, low self-efficacy, or high novelty seeking?
  \item Which AI design features increase or reduce the loop: conversational memory, unlimited responses, multi-model access, confidence markers, or source grounding?
  \item How does doom researching differ across domains such as academic writing, programming, health, finance, and career planning?
  \item Which interventions help users convert AI-assisted inquiry into knowledge and output?
\end{enumerate}

The most important empirical test is the perceived-actual knowledge gap. If AI-mediated repeated research reliably increases perceived knowledge more than actual knowledge or output, then doom researching would represent a meaningful human-AI interaction risk. If not, the term may still be useful descriptively, but its theoretical value would be weaker.

\section{Conclusion}

This paper proposed doom researching as a conceptual framework for understanding repetitive AI-assisted information seeking that fails to convert into proportional knowledge, decisions, or output. The construct draws from doomscrolling, information-seeking theory, cognitive offloading, transactive memory, metacognition, and human-AI interaction research. The proposed model suggests that AI systems can create a loop in which uncertainty reduction, fluent answers, reassurance, and cognitive offloading encourage continued querying, while synthesis and output are delayed. The result is a possible gap between feeling informed and being able to explain, decide, or produce.

Doom researching should not be understood as a rejection of generative AI. It is a call for better calibration. AI tools can support research when they help users ask better questions, verify claims, synthesize sources, and create artifacts. They become risky when asking replaces thinking, and when the feeling of knowing replaces knowledge itself.

\bibliographystyle{plainnat}
\bibliography{references}

\end{document}